\documentclass[twocolumn,showpacs,showkeys]{revtex4}
\usepackage{graphicx}
\usepackage{bm}
\usepackage{color}
\usepackage{amsmath}
\usepackage{natbib}

\begin{document}

\title{Vortical Dynamics of spinning quantum plasmas - Helicity Conservation}

\author{Swadesh M. Mahajan}
\email{mahajan@mail.utexas.edu}
\author{Felipe A. Asenjo}
\email{faz@physics.utexas.edu}
\affiliation{Institute for Fusion Studies, The University of Texas at Austin, Texas 78712, USA.}

\date{\today} 

\begin{abstract}
It is shown that a vorticity, constructed from spin field of a quantum spinning plasma, combines with the classical generalized vorticity (representing the magnetic and the velocity fields) to yield a new grand generalized vorticity that obeys the standard vortex dynamics. Expressions for the quantum or spin vorticity, and for the resulting generalized helicity invariant are derived. Reduction of the rather complex spinning quantum system to a well known and highly investigated classical form opens familiar channels for the delineation of physics peculiar to dense plasmas spanning solid state to astrophysical objects.  A simple example is worked out to show  that the magnetics of a spinning plasma can be much richer than that of the corresponding classical system.
\end{abstract}

\pacs{52.35.We, 67.10.-j, 67.25.dk, 67.30.hj}

\maketitle


In this paper we  demonstrate that a spinning quantum fluid plasma \cite{marklund,brodin} retains the most interesting and defining features of a classical ideal  fluid. We will show, in particular that  it is possible  to engineer a ``grand generalized vorticity'' (GGV) that obeys a vortex dynamic structure. Such a GGV is created by combining  the erstwhile ``generalized'' classical vorticity $\mathbf\Omega_c = \nabla\times\mathbf P_c$, where  $\mathbf P_c=\mathbf A+(mc/q)\mathbf v$ is proportional to the canonical momentum \cite{mah1,mah2}, and a ``quantum vorticity''  $\mathbf\Omega_q$ constructed from the macroscopic spin vector field $\mathbf S$. 

It is remarkable that we can rewrite  a complex and physically rich system as a quantum spinning plasma as a  standard vortex dynamics. At the very least it implies a new  composite constant of motion (the grand generalized helicity) and  the existence of an Alfven/Kelvin theorem. This formulation, however, has the potential for a far speedier extraction and exposition of a great many properties of spinning plasmas. The most important step in this new formulation is the construction/identification of the quantum vorticity vector $\mathbf\Omega_q$. As we will see, the form for $\mathbf\Omega_q$ is, by no means, obvious. Before embarking on the technical formulation, it is pertinent to put the current work in a historical perspective.

The ``project'' of the fluidization of quantum systems (Schrodinger, Pauli and Dirac equations) has been driven by two related  but distinct objectives:

1) Earlier investigators \cite{madelung,bohm,takabayasi2,takabayasi1}, wishing to understand and interpret quantum mechanics in terms of familiar classical concepts, were content to devise appropriate fluid-like variables obeying the ``expected'' fluid like equations of motion: for example  the continuity and  the  force balance equation. Quantum mechanics entered  the latter through the so called ``quantum forces'' proportional to powers of $\hbar$. The fluidized system, of course, was equivalent to the original quantum one.

2) After an extended hiatus following the initial studies in quantum plasmas \cite{pines, harris, dubois, balescu, gomberoff}, the impetus for the recent impressive comeback of the fluidization ``project'', however,  has come from a totally new direction-from attempts to investigate the  collective macroscopic motions accessible to a fluid (plasma) whose elementary constituents follow the laws of quantum rather than classical dynamics. The new chapter may be labelled, more appropriately,  as a macroscopic theory  of quantum plasmas as opposed to the earlier efforts that mostly consisted of casting quantum mechanics into a fluid-like mould. Much progress has been made in first constructing the desired macroscopic frameworks, and then working out and exposing new phenomena, originating in the quantum nature of the constituent particles. The macroscopic  formulations (for studying collective motions of quantum fluids) have invoked methodologies similar to those employed in classical plasmas;  both the fluid and kinetic theories of simple quantum \cite{feix,anderson,haas2,haas3,rus1}, spin quantum \cite{marklund,brodin,rus2, brodin2,jens,jens2,jens3,shuklaE}, and relativistic quantum plasmas \cite{asenjo,hakim,sivak,mendonca,biali} have been constructed.

The current work on the vortex dynamic formulation of spinning non relativistic quantum plasma, though highly influenced by Takabayasi's excellent papers spanning nineteen fifties to eighties \cite{takabayasi2,takabayasi1}, is of the latter genre for which the recent trend setting work of Marklund and Brodin \cite{marklund,brodin} provides a basic reference. The focus of this paper is on the elucidation of the basic concept of quantum vorticity. We will, therefore, work with the simplest model (the equivalent of an ideal classical fluid) obtained from Refs.~\cite{marklund,brodin} by neglecting complicated effects like the spin-spin and the thermal-spin couplings.

The spin quantum plasma is described by three coupled equations for the  density $n$, the fluid velocity $\mathbf v$ and the spin vector $\mathbf S$. First two are the continuity   
\begin{equation}
 \frac{\partial n}{\partial t}+\nabla\cdot(n\mathbf v)=0\, ,
\label{eccont}\end{equation}
and the momentum equation 
\begin{equation}
 m\left(\frac{\partial }{\partial t}+\mathbf v\cdot\nabla\right)\mathbf v=q\left(\mathbf E+\frac{\mathbf v}{c}\times\mathbf B\right)+\mu S^j\nabla {\widehat B}_j+\mathbf\Xi\, ,
\label{ecmomentum}\end{equation}
with 
\begin{equation}
{\mathbf {\widehat B}}=\mathbf B+\frac{\hbar c}{2qn}\partial^i\left(n\partial_i \mathbf S\right)\, , 
\end{equation}
where $q$ ($m$) is the particle charge (mass), $\mathbf E$ and $\mathbf B$ are the electric and magnetic field respectively, $\mu=q\hbar/2mc$ is the elementary magnetic moment, $\hbar$ is the reduced Planck constant, $c$ is the speed of light, $S_j$ is the $j$-component for the normalized unit-modulous spin vector $\mathbf S$ ($\mathbf S\cdot\mathbf S=1$), and ${\widehat B}_j$ is the $j$-component of $\mathbf{\widehat B}$. Notice, that the usual spin vector used in Refs.~\cite{marklund,brodin} is $\hbar \mathbf S/2$. 

The last term in the momentum equation is the force produced by the total fluid pressure  
\begin{equation}
\mathbf\Xi=-\frac{1}{n}\nabla p+\frac{\hbar^2}{2m}\nabla\left(\frac{\nabla^2\sqrt{n}}{\sqrt{n}}\right)+\frac{\hbar^2}{8m}\nabla\left(\partial^jS^i\partial_jS_i\right)\, ,
\end{equation}
consisting of the classical pressure $p$, the Bohm potential (the second term), and the effective spin pressure.

The third equation is the evolution of spin vector
\begin{equation}
 \left(\frac{\partial }{\partial t}+\mathbf v\cdot\nabla\right)\mathbf S=\frac{2\mu}{\hbar}\left(\mathbf S\times\mathbf{\widehat B}\right)\, ,
\label{ecspin}\end{equation}
that is similar to the classical preccesion equation for the spin with the spin correction to the magnetic field. The set of Eqs.~\eqref{eccont}, \eqref{ecmomentum} and \eqref{ecspin} is completely equivalent to those found in the primary Refs.~\cite{marklund,brodin}. 

Let us now convert the system into evolution equations for the appropriately defined vorticities. Using
 $\mathbf E=-\nabla\phi-\partial_t\mathbf A/c$, $\mathbf B=\nabla\times\mathbf A$ ($\phi$ and $\mathbf A$ are the scalar and vector potentials), and the vector identity $\left(\mathbf v\cdot\nabla\right)\mathbf v = \nabla\mathbf{v}^2/2 -\mathbf v\times({\nabla\times\mathbf v})$, Eq.~\eqref{ecmomentum} becomes
\begin{equation}
 \frac{\partial \mathbf P_c}{\partial t}=\mathbf v\times \mathbf\Omega_c+\frac{\hbar}{2m}S^j\nabla{\widehat B}_j+\frac{c}{q}{\widehat \mathbf\Xi}\, ,
\label{ecmomentumpotential}\end{equation}
where ${{\widehat \mathbf\Xi}}= \mathbf\Xi-\nabla(q\phi+m\mathbf{v}^2/2)$, and  $\mathbf P_c$ is proportional to the classical canonical momentum
\begin{equation}
 \mathbf P_c=\mathbf A+\frac{mc}{q}\mathbf v\, .
\end{equation}
The ensuing classical generalized vorticity (vorticity will have the dimensions of the magnetic field throughout this paper) 
\begin{equation}
\mathbf\Omega_c= \nabla\times\mathbf P_c=\mathbf B+\frac{mc}{q}\nabla\times\mathbf v\, ,
\end{equation}
will, then, obey 
\begin{equation}
 \frac{\partial \mathbf \Omega_c}{\partial t}=\nabla\times\left(\mathbf v\times \mathbf\Omega_c\right)+\frac{\hbar}{2m}\nabla S^j\times\nabla {\widehat B}_j\, ,
\label{clasvortic}\end{equation}
obtained, by taking the the curl of Eq.~\eqref{ecmomentumpotential} and  having assumed a barotropic fluid. We notice that the spin dependent forces destroy the canonical vortical structure for $\mathbf \Omega_c$ \cite{mah1,mah2}. Consequently, the classical generalized helicity [$\langle\, \rangle=\int d^3x$]
\begin{equation}
h_c=\langle\mathbf \Omega_c\cdot\mathbf P_c\rangle\, , 
\end{equation}
is no longer conserved. We remind the reader that the (generalized) helicity conservation is one of the most important properties of ideal fluids and is the primary dynamical constraint that allows the formation of a host of non trivial self-organizing equilibrium configurations in magnetohydrodynamics, and also in more general plasma descriptions. The loss of a helicity invariant could make it much harder to understand the fundamental motions of a spinning quantum fluid 

One could take the alternative view that spin forces act as a quantum source (proportional to $\hbar$) that may create or destroy  
helicity via
\begin{equation}
 \frac{d h_c}{d t}=\frac{\hbar}{m}\left\langle{\Omega_c}^i S^j\partial_i {\widehat B}_j\right\rangle\, ,
\end{equation}
and, in the process, cause transitions to a different helicity state. 
Observe that only the spin force, being non potential, survives in the vortical equation. The potential quantum forces like the Bohm potential do not contribute to the vorticity evolution.

Experience, however, indicates that, though, addition of new physics (to fluid mechanics) does destroy old invariants, new and more encompassing new invariants often emerge \cite{bek,mah2003}. Spinning quantum plasmas prove to be no exception! Guided by Takabayasi's  work \cite{takabayasi1}, we were able to uncover, what could be called, the spin or quantum vorticity:
\begin{eqnarray}
\mathbf\Omega_q&=&S_1\left(\nabla S_2\times \nabla S_3\right)+S_2\left(\nabla S_3\times \nabla S_1\right)\nonumber\\
&&+S_3\left(\nabla S_1\times \nabla S_2\right)=\left(\nabla S_1\times \nabla S_2\right)/ S_3\, ,
\label{vorticitq1}
\end{eqnarray}
where the components of $\mathbf S$ are labeled by $1,2,3$. Equality of the two expressions, displayed in  Eq.~\eqref {vorticitq1}, follows from the constraint  $S_1^2+S_2^2+S_3^2=1$ implying $S_1 \nabla S_1+S_2\nabla S_2+S_3 \nabla S_3=0$. For completeness, the quantum vorticity could be also written in the component form as $ {\Omega_q}^i= (1/2)\varepsilon^{ijk}\varepsilon^{lmn}S_l\partial_j S_m \partial_k S_n $. 

The quantum vorticity associated with the spin field has many interesting features. First, it requires that all $S_i$ and $\nabla S_i$ to be non zero; the system must have variation in at least two dimensions for a non trivial  $\mathbf\Omega_q$. Secondly, although symmetric in the three spin components, its form could not be easily guessed; it departs so fundamentally from the form taken by the vorticity $\nabla\times\mathbf v$ (or $\nabla\times\mathbf A$) associated with the standard classical vector fields $\mathbf v$ (or $\mathbf A$). In spite of these peculiarities, it does conform to our notions of a vorticity, i.e, it is the curl of a vector field: $\mathbf\Omega_q$=$\nabla\times\mathbf P_q$, with $\mathbf P_q= -S_3 \nabla[ \arctan(S_2/S_1)]$. The vector field is in the Clebsch form.

Manipulations of the spin dynamical equation \eqref{ecspin} yields the evolution equation 
\begin{equation}
 \frac{\partial \mathbf \Omega_q}{\partial t}=\nabla\times\left(\mathbf v\times \mathbf\Omega_q\right)+\frac{q}{mc}\nabla S^j\times\nabla {\widehat B}_j\, ,
\label{quantumvortic}
\end{equation}
and its uncurled companion for the potential $\mathbf P_q$ \cite{takabayasi1}
\begin{equation}
 \frac{\partial \mathbf P_q}{\partial t}=\mathbf v\times \mathbf \Omega_q+\frac{q}{mc}S^j\nabla {\widehat B}_j\, .
\label{quantumvorticpotential}
\end{equation}
Notice that $\mathbf\Omega_q$ obeys exactly the same equation \eqref{clasvortic} as is obeyed by $\mathbf\Omega_c$. This is, of course, no accident; it was the entire raison d'etre for constructing $\mathbf\Omega_q$.
 The journey from \eqref{ecspin} to \eqref{quantumvortic} is both unusual and profound.

By adding and subtracting Eqs.\eqref{clasvortic} and \eqref{quantumvortic}, we derive the two  GGVs, $\mathbf\Omega_+$ and $\mathbf\Omega_-$
\begin{equation}
 \mathbf\Omega_{\pm}=\mathbf\Omega_c\pm\frac{\hbar c}{2q}\mathbf \Omega_q\, ,
\end{equation}
explicitly showing that quantum modification to the classical vortex field is of order $\hbar/2$. The new vorticities follow: 
\begin{equation}
\frac{\partial \mathbf \Omega_+}{\partial t}=\nabla\times\left(\mathbf v\times \mathbf\Omega_+\right)+\frac{\hbar}{m}\nabla S^j\times\nabla {\widehat B}_j\, ,
\label{vorticplus}
\end{equation}
\begin{equation}
\frac{\partial \mathbf \Omega_-}{\partial t}=\nabla\times\left(\mathbf v\times \mathbf\Omega_-\right)\, .
\label{vorticminus}
\end{equation}
while the associated potential vector fields $\mathbf P_{\pm}=\mathbf P_c\pm (\hbar c/2q)\mathbf P_q$ satisfy
\begin{equation}
 \frac{\partial \mathbf P_+}{\partial t}=\mathbf v\times \mathbf\Omega_++\frac{\hbar}{m}S^j\nabla {\widehat B}_j+\frac{c}{q}{{\widehat \mathbf\Xi}}\, ,
\label{potentialmas}\end{equation}
 \begin{equation}
 \frac{\partial \mathbf P_-}{\partial t}=\mathbf v\times \mathbf\Omega_-+\frac{c}{q} {{\widehat \mathbf\Xi}}\, ,
\label{potentialmenos}\end{equation}

Equation \eqref{vorticminus} is clearly what we were seeking; the grand generalized vorticity  $\mathbf\Omega_-$ obeying the  canonical vortex dynamics.  Thus the structure of the dynamics of a spinning quantum plasma, in part, has been reduced to that of a highly investigated and understood classical system. The conserved helicity $h_-=\langle\mathbf P_-\cdot\mathbf\Omega_-\rangle$,    
\begin{equation}
 \frac{d h_-}{d t}=0\, .
\end{equation}
will serve as a ``label'' to characterize  dynamical states of a spinning quantum plasma.

It turns out, however, that the highly complex spinning quantum plasma demands a two vorticity theory with only one of them as a basic invariant. The second generalized quantum helicity $h_+=\langle\mathbf P_+\cdot\mathbf\Omega_+\rangle$ is not conserved, and its rate of change is given by
\begin{equation}
 \frac{d h_+}{d t}=\frac{2\hbar}{m}\left\langle{\Omega_+}^i S^j\partial_i {\widehat B}_j\right\rangle\, .
\end{equation}
In the wake of Eqs.\eqref{clasvortic} and \eqref{quantumvortic}, the rate of change of either $h_c$ or $h_q$ is proportional to $dh_+/dt$. 

To the vorticity equations, we add Maxwell's equations 
\begin{equation}
 \nabla\times\mathbf B=\frac{4\pi}{c}\mathbf J+4\pi\nabla\times\mathbf M+\frac{1}{c}\frac{\partial \mathbf E}{\partial t}\, ,
\label{maxwell}
\end{equation}
to complete the dynamical system consisting of the magnetic, velocity and spin fields. It contains the normal current density
 $\mathbf J$, and $\mathbf M=\mu n\mathbf S $ is the magnetization that defines the spin current density $\nabla\times\mathbf M$ \cite{brodin}.  
 
The main intent of this paper was  to create the conceptual foundation for the vortex dynamic formulation of a spinning quantum plasma. The next obvious step will be to explore the class of equilibrium structures pertinent to a spinning plasma by invoking the constrained (conserving $\Omega_-$) minimization of an appropriate energy functional \cite{mah2}. We will defer this investigation to a later detailed paper and solve here a simple equilibrium problem that may be viewed as a generalization of the London equation, first proposed, to explain the Meissner-Ochsefeld effect observed in type one superconductors. Electrodynamically, the London equation is nothing but the absence of generalized vorticity \cite{mah3}
\begin{equation}
\mathbf\Omega_c= \mathbf B+\frac{mc}{q}\nabla\times\mathbf v=0 .
\label{London}
\end{equation}
Combined with the displacement current-free maxwell equation it  yields the strongly diamagnetic behavior where the magnetic field ($\lambda_s^2\nabla^2\mathbf B=\mathbf B$) is limited to a skin depth  $\lambda_s=c/\omega_p$ (where $\omega_p=(4\pi q^2 n/m)^{1/2}$ is the plasma frequency) near the edge of a region of length $L$ ($\gg\lambda_s$). 

The generalization of the London equation for the  spin quantum system,   
$\mathbf\Omega_-=0$, will span new equilibrium structures.  
This class of such equilibria, defined by the vorticity equations \eqref{vorticplus}, \eqref{vorticminus} and Maxwell equations \eqref{maxwell}, for an incompressible fluid ($\nabla\cdot{\mathbf v}=0$) with  constant number density,  may be converted to the dimensionless  set:
\begin{equation}
{\nabla}\times\left({\mathbf\Omega}_q\times{\mathbf v}\right)=\nabla S^j\times\nabla\left(b_j+a\nabla^2S_j\right)\, ,
\label{normalized1}\end{equation}
\begin{equation}
 \mathbf b+\nabla\times{\mathbf v}=a{\mathbf\Omega}_q\, ,
\label{normalized2}\end{equation}
\begin{equation}
\nabla\times\mathbf b={\mathbf v}+a\nabla\times\mathbf S\, ,
\label{normalized3}
\end{equation}
with the following normalizations: all lengths to $\lambda_s^{-1}$, magnetic field to a fiducial field $B$, and velocity to the Alfven speed $v_A=c \omega_c/\omega_p$, where $\omega_c=qB/mc$ is the cyclotron frequency associated with the magnetic field.

Remarkably enough, the entire system has a single characteristic parameter $a=\lambda_c\omega_p/(2 v_A)=(\lambda_c^2/\lambda_s^2)(mc^2/\hbar\omega_c)$ that determines the relative strength of the newly found quantum vorticity to the canonical vorticity. It may be viewed as the ratio between the Compton length $\lambda_c=\hbar/mc$ and the classical length $v_A/ \omega_p$. It could also be viewed as the square of the ratio $\lambda_c/\lambda_s$ ehanced by the ratio between the particle rest mass and the ``quantized magnetic energy''. The quantum  contribution tends to become more and more significant as the density increases and as the magnetic field decreases.

For simplicity we assume a two dimensional variation with $ \partial /\partial z = 0$ and $\nabla=\hat e_xd/dx+\hat e_yd/dy$. For the spin vector $\mathbf S$, we propose the solution: $\mathbf S(x,y)=\hat e_x~g(x)\cos y+\hat e_y~g(x)\sin y+\hat e_z~f(x)$, such that $f^2+g^2=1$. For this ansatz, only the $\hat e_z$ component survives for the spin vorticity, $\mathbf\Omega_q=-\hat e_z f'(x)$, where $' = d/dx$.  

The inherent symmetry  of the system suggests the following form for the magnetic field: $\mathbf b=\hat e_x~p_1(x)\cos y+\hat e_y~p_2(x)\sin y+\hat e_z~p_3(x)$. For these forms of $\mathbf S$ and $\mathbf B$, the equilibrium set reduces to ordinary differential equations in $x$.  Equations~\eqref{normalized2} and \eqref{normalized3} yield
\begin{equation}
 2 p_1-p''_1=a\left(g'+g\right)\, ,
\label{ecsolexactp1}
\end{equation}
\begin{equation}
 2 p_2-p''_2=-a\left(g''+g'\right)\, ,
\label{ecsolexactp2}\end{equation}
\begin{equation}
 p''_3-p_3=a\left(f''+f'\right)\, ,
\label{ecsolexactp3}\end{equation}
out of which \eqref{ecsolexactp2} collapses to \eqref{ecsolexactp1} because $\nabla\cdot\mathbf b=(p_1'+p_2)\cos y=0$ (or $p_2=-p_1'$).

 The set \eqref{ecsolexactp1}-\eqref{ecsolexactp3} is augmented by a third equation derived from Eq.~\eqref{normalized1} whose left hand side is identically zero and the right hand side has only $\hat e_z$-component. The third equation $g(p_1''+p_1')=g'(p_1'+p_1)$   integrates to
\begin{equation}
 p_1'+p_1=\alpha g\, ,
\label{eqexact32}
\end{equation}
where $\alpha$ is a constant which must be determined by boundary conditions. The fields $\mathbf b$ and $\mathbf S$ will be known when we solve Eqs.~\eqref{ecsolexactp1}, \eqref{ecsolexactp3} and \eqref{eqexact32}.
In analogy with a superconducting solution, let us consider a domain $0<x<L$ (with periodic behavior in $y$), with  $L\gg 1$ (normalized to the skin depth). It is straightforward to verify that  a consistent solution to the whole system is:  $g(x)=e^{k(x-L)}$ ($g\leq 1$), $p_1=\alpha e^{k(x-L)}/(k+1)$, and
\begin{equation}
p_3(x)=e^{x-L}+a e^x \int_L^xdx' e^{x'}\frac{d}{dx'}\sqrt{1-e^{2k(x'-L)}}\, ,
\end{equation}
where we have used $f=\sqrt{1-g^2}$. The scale factor $k=\left(-a_q\pm\sqrt{2+a_q}\right)/(1+a_q)$, where ${a}_q=a/\alpha$.

Remembering that  the classical solution is normally taken to be  $b_1=0=b_2$, and  $b_3= e^{x-L}$ (extreme diamagnetism ($L\gg1$) with field non zero only in a skin depth), we find that the spin field has transformed it fundamentally: 1) The field $b_3= p_3$ in the spinning plasma has an additional quantum contribution proportional to $a$ with a new "quantum scale" $k$. 2) Magnetic field components perpendicular to  spin vorticity, $b_1=p_1\cos y$, and $b_2=-{p_1}'\sin y$, emerge; their magnitude is proportional to the spin vorticity.
Detailed discussion and implications of this particular solution, and also of other solutions, including the ones in which the  quantum spin vorticity may dominate  its classical counterparts, will be given in a future paper. The main objective of this paper was to
construct an appropriate spin/quantum vorticity that will lead to the emergence of a new generalized quantum vorticity $\mathbf\Omega_-$ obeying the standard vortex dynamics of the Helmholz form.  Finding $\mathbf\Omega_-$ that guarantees  the existence of a dynamical helicity invariant, constitutes the main mathematical results of this paper. It is hoped that the vortex dynamic structure will greatly aid in extracting new physics inherent in the spinning plasmas.


The work of SMM was supported by USDOE Contract No.DE-- FG 03-96ER-54366. FAA thanks the CONICyT for a BecasChile Postdoctoral Fellowship.


\begin{thebibliography}{17}

\bibitem{marklund} M. Marklund and G. Brodin, Phys. Rev. Lett. {\bf 98}, 025001 (2007).

\bibitem{brodin} G. Brodin and M. Marklund, New J. Phys. {\bf 9}, 277 (2007).

\bibitem{mah1} S. M. Mahajan, Phys. Rev. Lett. {\bf 100}, 075001 (2008).

\bibitem{mah2} S. M. Mahajan and Z. Yoshida, Phys. Rev. Lett. {\bf 81}, 4863 (1998).

\bibitem{madelung} E. Madelung, Zeit. F. Phys. {\bf 40}, 322 (1927).

\bibitem{bohm} D. Bohm, Phys. Rev. {\bf 85} 166 (1952).

\bibitem{takabayasi2} T. Takabayasi,  Prog. Theo. Phys. {\bf 14}, 283 (1955); Prog. Theo. Phys. {\bf 12}, 810 (1954); Prog. Theo. Phys. {\bf 13}, 222 (1955); Phys. Rev. {\bf 102}, 297 (1956); Nuovo Cimento {\bf 3}, 233 (1956); Prog. Theo. Phys. Suppl. {\bf 4}, 2 (1957).

\bibitem{takabayasi1} T. Takabayasi, Prog. Theor. Phys. {\bf 70}, 1 (1983).

\bibitem{pines} D. Pines, J. Nucl. Energy, Part C {\bf 2}, 5 (1961).	

\bibitem{harris} E. G. Harris, Advances in Plasma Physics {\bf 3}, 157, ed. by A. Simon and W. B. Thompson, Interscience Publ.(1969).

\bibitem{dubois} D. F. DuBois, Lectures in Theoretical Physics  {\bf
    9C}, 469,  W. E. Brittin and A. D. Barut Editors, Gordon-Breach, New York (1967).

\bibitem{balescu} R. Balescu, Phys. Fluids {\bf 4}, 94 (1961).

\bibitem{gomberoff} L. Gomberoff, J. Plasma Phys. {\bf 18}, 145 (1977).

\bibitem{feix} F. Haas, G. Manfredi and M. Feix, Phys. Rev. E {\bf 62}, 2763 (2000).

\bibitem{anderson} D. Anderson, B. Hall, M. Lisak, and M. Marklund, Phys. Rev. E {\bf 65}, 046417 (2002).

\bibitem{haas2} F. Haas, Phys. Plasmas {\bf 12}, 062117 (2005).

\bibitem{haas3} F. Hass {\it et al.}, Phys. Lett. A {\bf 374}, 481 (2010).

\bibitem{rus1} L. S. Kuz'menkov and S. G . Maksimov, Theor. Math. Phys. {\bf 118}, 227 (1999).

\bibitem{rus2} P. A. Andreev and L. S. Kuz'menkov, Russ. Phys. Jour. {\bf 50}, 1251 (2007).


\bibitem{brodin2} G. Brodin {\it et al.}, Phys. Rev. Lett. {\bf 101}, 245002 (2008)

\bibitem{jens} J. Zamanian {\it et al.}, New J. Phys. {\bf 12}, 043019 (2010).

\bibitem{jens2} F Haas {\it et al.} New J. Phys. {\bf 12}, 073027 (2010).

\bibitem{jens3} J. Zamanian {\it et al.}, Phys. Plasmas {\bf 17}, 102109 (2010).

\bibitem{shuklaE} P. K. Shukla and B. Eliasson, Phys.-Uspekhi {\bf 53}, 51 (2010).

\bibitem{asenjo} F. A. Asenjo {\it et al.}, Phys. Plasmas {\bf 18}, 012107 (2011).

\bibitem{hakim} R. Hakim and J. Heyvaerts, Phys. Rev. A {\bf 18}, 1250 (1978).

\bibitem{sivak} H. D. Sivak, Ann. Phys {\bf 159}, 351 (1985).

\bibitem{mendonca} J. T. Mendon\c{c}a, Phys. Plasmas {\bf 18}, 062101 (2011).

\bibitem{biali} I. Bialynicki-Birula, P. G\'ornicki and J. Rafelski,  Phys. Rev. D {\bf 44}, 1825 (1991).  

\bibitem{bek} J. D. Bekenstein, Astrophys. J. {\bf 319}, 207 (1987).

\bibitem{mah2003} S. M. Mahajan, Phys. Rev. Lett. {\bf 90}, 035001 (2003)

\bibitem{mah3} S. M. Mahajan, Phys. Rev. Lett. {\bf 100}, 075001 (2008)











\end{thebibliography}
\end{document}